\documentclass[11pt,a4paper]{article}
\usepackage{jheppub}
\usepackage{amsmath,amssymb,epsf}
\usepackage{graphicx,color}

\def\be{\begin{equation}}
\def\ee{\end{equation}}
\def\bea{\begin{eqnarray}}
\def\eea{\end{eqnarray}}

\def\ltap{\ \raise.3ex\hbox{$<$\kern-.75em\lower1ex\hbox{$\sim$}}\ }
\def\gtap{\ \raise.3ex\hbox{$>$\kern-.75em\lower1ex\hbox{$\sim$}}\ }
\def\gl{\ \raise.5ex\hbox{$>$}\kern-.8em\lower.5ex\hbox{$<$}\ }
\def\roughly#1{\raise.3ex\hbox{$#1$\kern-.75em\lower1ex\hbox{$\sim$}}}

\def\inf{\text{inf}}
\def\mod{\text{mod}}
\def\up{\text{up}}
\def\AdS{\text{AdS}}
\def\mix{\text{mix}}
\def\tot{\text{tot}}
\def\Hinf{\mathcal{H}_{\inf}}

\newcounter{oldcounter}
\addtocounter{equation}{1}
\title{Combining High-scale Inflation with Low-energy SUSY}
\author[a, b]{Stefan Antusch,}
\author[c]{Koushik Dutta,}
\author[b]{Sebastian Halter}
\affiliation[a]{
Department of Physics, University of Basel, \\
Klingelbergstr. 82, CH-4056 Basel, Switzerland}
\affiliation[b]{Max-Planck-Institut f\"ur Physik (Werner-Heisenberg-Institut)\\ 
F\"ohringer Ring 6, D-80805 M\"unchen, Germany}
\affiliation[c]{DESY, Theory Group,\\Notkestrasse 85, D-22607 Hamburg, Germany }
\emailAdd{stefan.antusch@unibas.ch}
\emailAdd{koushik.dutta@desy.de}
\emailAdd{halter@mppmu.mpg.de}
\abstract{
We propose a general scenario for moduli stabilization where low-energy supersymmetry can be 
accommodated with a high scale of inflation. The key ingredient is that the stabilization
of the modulus field \emph{during} and \emph{after} inflation is not associated
with a single, common scale, but relies on two different mechanisms. We illustrate this general scenario in a simple example, where during inflation the modulus is stabilized with a large mass by a K\"ahler potential coupling to the field which provides the inflationary vacuum energy via its F-term. After inflation, the modulus is stabilized, for instance, by a KKLT superpotential. 
}
\keywords{Inflation, Supersymmetry Breaking, Moduli Stabilization}
\arxivnumber{1112.4488}

\begin{document}

\begin{flushleft}
MPP-2011-153 \hfill DESY  11-250
\end{flushleft}

\maketitle

\section{Introduction}
\label{Sec:Introduction}

Low-energy supersymmetry (SUSY) at the TeV-scale provides an attractive extension of the Standard Model (SM) for various reasons, in particular, since it can solve the gauge hierarchy problem. Furthermore, it is going to be tested in the ongoing LHC experiments. Regarding cosmology, the paradigm of cosmic inflation has emerged as the prime candidate for early universe physics to resolve the flatness and horizon problems associated with the hot big bang scenario. Typically, models of inflation require a comparatively high scale of inflation, close to the energies where the gauge interactions of the SM can be unified. The scale of inflation might soon be tested by an observation of tensor modes with the PLANCK satellite. It is therefore an interesting question whether a high scale of inflation can be realized together with TeV-scale SUSY.

A priori, the two topics of low-energy SUSY and high-scale inflation seem to be unrelated. However, they turn out to be connected due to the issue of ``moduli stabilization'', in particular in the context of string theory. String compactifications generically lead to many light scalar fields, aka moduli, which for example parametrize the size and shape of the compactified internal manifolds. Tracing the physics of extradimensions, the couplings in the 4D effective low-energy theory become functions of the moduli fields. To make sure that the internal dimensions do not decompactify, and also to avoid observational constraints on the space-time variability of the low-energy coupling constants, the moduli fields must be stabilized (i.e. must acquire a suitable mass) both during and after inflation. 

In the context of type IIB string theory, the issue of moduli stabilization is most well understood in the well-known KKLT scenario \cite{Kachru:2003aw}. It assumes that the dilaton and the complex structure moduli are stabilized via fluxes \cite{Giddings:2001yu,Dasgupta:1999ss,Taylor:1999ii}.\footnote{For reviews and an extensive list of references on moduli stabilization with fluxes and nonperturbative effects in type IIB string theory cf. e.g. \cite{Grana:2005jc,Douglas:2006es,Blumenhagen:2006ci,Denef:2007pq}.} Therefore, only the dynamics of the volume moduli are important for the low-energy physics. The volume moduli are stabilized by the contributions from nonperturbative effects such as gaugino condensates. Moreover, moduli stabilization also nicely combines with dynamical supersymmetry breaking with the moduli comprising the hidden sector.

When the issue of moduli stabilization is discussed in conjunction with inflation, however, a severe problem emerges, namely adding the inflationary sector may destabilize the moduli, which was pointed out by Buchm\"uller, Hamaguchi, Lebedev and Ratz and by Kallosh and Linde in \cite{Buchmuller:2004xr,Buchmuller:2004tz,Kallosh:2004yh}. Here, we are concerned with a particular version of this problem sometimes referred to as the Kallosh-Linde (KL) problem \cite{Kallosh:2004yh}: One often finds an upper bound on the inflationary energy scale in terms of the present-day gravitino mass,
\begin{equation}
\Hinf \leq m_{3/2}^{\text{today}} \, ,
\end{equation}
to avoid the destabilization of the volume modulus. For TeV-scale SUSY breaking, one has $m_{3/2} \sim \text{TeV}$ and thus the scale of inflation is bound to be very small, much below the scale required for
many model building approaches (and also very much below observational
sensitivities for gravitational waves). Basically, the problem appears since
there is effectively only one scale in the problem, which sets both the
gravitino mass today and the height of the barrier towards decompactification. A
SUSY breaking uplifting term turns the AdS minimum into a nearly Minkowski
minimum, and also sets the height of the barrier separating the metastable
vacuum from the vacuum at infinity. Typically, inflation in such a setup can be
viewed as an additional uplifting which induces a runaway potential for the
modulus. Thus, if its contribution becomes too large, the barrier and hence the
minimum disappear. In summary, the inflationary scale is constrained to be
smaller than the height of the barrier, and therefore the gravitino mass. We
will review the problem in more detail in the next section.

As a solution to the problem, KL \cite{Kallosh:2004yh} suggested a form of the superpotential for which the gravitino mass in the present vacuum is unrelated to the height of the barrier. Therefore, choosing the gravitino mass at the TeV-scale, we can always independently increase the barrier height such that high-scale inflation models do not destabilize the modulus. This corresponds to a fine-tuning of the terms in the superpotential. Following the approach of KL, in the context of volume modulus inflation in a racetrack setup, the problem has been addressed when one of the exponents of the nonperturbative terms is positive \cite{Abe:2005rx,Badziak:2008yg,Abe:2008xu,Badziak:2008gv,Badziak:2009eh}. However, this can be done only at the expense of introducing more parameters in the theory. In the large volume scenario, attempts have been made to accommodate a small gravitino mass with a high scale of inflation, when inflation happens exponentially far away from the present Minkowski vacua. Other than some inevitable fine tuning, the working models have several phenomenological difficulties \cite{Conlon:2008cj}. Dynamical avenues  for the case of chaotic inflation \cite{He:2010uk} and hybrid inflation \cite{Kobayashi:2010rx} have also been explored, where the gravitino mass becomes inflaton-dependent in a suitable way. Difficulties related to the realizations of high-scale inflation together with low-energy SUSY breaking have been discussed in \cite{Davis:2008sa,Davis:2008fv} and some resolutions have been proposed in \cite{Mooij:2010cs,Davis:2008sa}. However, this has been successfully achieved only for the superpotential of \cite{Kallosh:2004yh}. Combining chaotic inflation and supersymmetry breaking within the KL scheme has recently been discussed in \cite{Kallosh:2011qk} for the general chaotic inflation models of \cite{Kallosh:2010ug,Kallosh:2010xz,Demozzi:2010aj}. For other approaches to the KL problem see e.g. \cite{Misra:2009ei}.

Here, we propose a different resolution of the problem, namely to use two different mechanisms to stabilize the moduli during and after inflation. During inflation, the modulus field receives a large mass proportional to the inflationary vacuum energy. This is achieved, for example, by a suitable moduli-dependence of the K\"ahler metric of the field driving supersymmetry breaking during inflation. The role of such a type of coupling for moduli stabilization was first noted in \cite{Antusch:2008pn}. At the end of inflation, the vacuum energy goes away and we invoke a stabilization mechanism like KKLT relying as usual on nonperturbative terms in the superpotential. We illustrate our general idea in a simple example: chaotic inflation protected by a shift symmetry combined with a KKLT-type superpotential. However, we stress that our framework can be applied to more general setups.

We begin with a brief review of the Kallosh-Linde Problem in section~\ref{Sec:ReviewKalloshLindeProblem}. Afterwards, we describe our general framework for resolving the problem in section~\ref{Sec:ResolutionGeneralFramework} and illustrate it in a simple example with shift symmetric chaotic inflation and a KKLT superpotential in section~\ref{Sec:AnExplicitExample}. Some results for a generic choice of $W_{\mod}(T)$ are presented in appendix \ref{Sec:AppendixGeneralWmod}. Finally, we give our conclusions in section~\ref{Sec:Conclusion}.

\section{Review of the Kallosh-Linde Problem}
\label{Sec:ReviewKalloshLindeProblem}

The Kallosh-Linde (KL) problem was discussed in \cite{Kallosh:2004yh} in the context of moduli stabilization in type IIB string theory within the KKLT scenario \cite{Kachru:2003aw}. Below the scale where the complex structure moduli and the dilaton are stabilized via fluxes as in \cite{Giddings:2001yu,Dasgupta:1999ss}, the K\"ahler potential $K$ and the superpotential $W$ for the volume modulus $T \equiv \sigma + i \, \alpha$, which controls the overall size of the compact space, are given by \cite{Kachru:2003aw}
\begin{equation}
  K = - 3 \ln (T +\bar{T}) \,\,\,\, \text{and} \,\,\,\, W = w_{0} + A e^{- a \, T} \, ,
\end{equation}
respectively. Note that throughout this article we work in units where $M_{P} \equiv 1$. In the following, we consistently set the imaginary part $\alpha = 0$, which corresponds to a particular choice of phases for $w_{0}$ and $A$, and consider only the potential for the real part $\sigma$. The resulting F-term potential has a supersymmetric Anti de Sitter (AdS) minimum and consequently the depth of this minimum is given by
\begin{equation}
  \label{Eq:KLProblemAdSMinimumDepth}
  V_{\AdS} = - 3 \, e^{\langle K \rangle_{\sigma_{\AdS}}} \, \lvert \langle W \rangle_{\sigma_{\AdS}} \rvert^{2} \, ,
\end{equation}
where $\sigma_{\AdS}$ is the position of the AdS minimum. To turn the this into a Minkowski or de Sitter (dS) minimum, one has to add an uplifting term to the potential, which is typically of the form $\Delta V \sim \frac{c_{\up}}{\sigma^{2}}$.\footnote{This choice for $V_{\up}$ is motivated by introducing $\overline{D3}$-branes at the tip of a warped throat. The constant $c_{\up}$ is tuneable by adjusting the strength of the warping at the tip.} The value of $c_{\up}$ is fine-tuned such that the value of the potential at the new minimum is equal to the present value of the cosmological constant. The uplifting usually induces only a small shift in the position of the minimum which is negligible, i.e. one has $\sigma_{0} \approx \sigma_{\AdS}$. The uplifting procedure also creates a barrier that prevents the field from running away to the Minkowski vacuum at $\sigma \rightarrow \infty$, i.e. towards  decompactification. The height $V_{B}$ of this barrier turns out to be $V_{B} \simeq \mathcal{O}(1) \lvert V_{\AdS} \rvert$.

The gravitino mass in the uplifted minimum is given by
\begin{equation}
  \label{Eq:KLProblemGravitinoMassUplift}
  m_{3/2}^{2}(\sigma_{0}) = e^{\langle K \rangle_{\sigma_{0}}} \lvert \langle W \rangle_{\sigma_{0}} \rvert \approx e^{\langle K \rangle_{\sigma_{\AdS}}} \lvert \langle W \rangle_{\sigma_{\AdS}} \rvert = \frac{1}{3} \lvert V_{\AdS} \rvert \, .
\end{equation}
Thus, the height of the barrier is related to the gravitino mass in the present vacuum,
\begin{equation}
  \label{Eq:KLProblemRelationGravitinoMassBarrierHeight}
  V_{B} \sim \lvert V_{\AdS} \rvert \sim m_{3/2}^{2} \, .
\end{equation}
The gravitino mass $m_{3/2}$ is directly related to the scale of supersymmetry breaking since the almost vanishing of the cosmological constant,
\begin{equation}
  \label{Eq:KLProblemRelationGravitinoMassSUSYBreaking1}
  V_{\mod} = V_{F} + V_{D} = \lvert F \rvert^{2} - 3 m_{3/2}^{2} + \frac{1}{2} D^{2} \approx 0 \, ,
\end{equation}
automatically implies
\begin{equation}
  \label{Eq:KLProblemRelationGravitinoMassSUSYBreaking2}
  3 m_{3/2}^{2} \approx \lvert F \rvert^{2} + \frac{1}{2} D^{2} \, .
\end{equation}

When an inflationary sector is added to the moduli stabilizing sector, the generic form of the potential becomes\footnote{If we add an F-term driving inflation, the scalar potential during inflation is
\begin{equation}
  \nonumber
  V = e^{K} \, \left( K^{X \bar{X}} \, \lvert D_{X} W \rvert^{2} + K^{T \bar{T}} \, \lvert D_{T} W \rvert^{2} - 3 \lvert W \rvert^{2} \right) + \frac{c_{\up}}{\sigma^{2}} \, .
\end{equation}

For all known candidate inflation models, the inflationary potential $V_{\inf} \sim e^{K} \, K^{X \bar{X}} \, \lvert D_{X} W \rvert^{2}$ vanishes as some inverse power of $\sigma$ for $\sigma \rightarrow \infty$. Typically, this power is $\sigma^{-3}$ from the $e^{K}$ prefactor. Thus, adding inflation to $V_{\mod}$ can be viewed as an additional uplifting.
}
\begin{equation}
  \label{Eq:KLProblemVmodPlusVinf}
  V_{\tot} = V_{\mod}(\sigma) + \frac{V_{\inf}(\phi)}{\sigma^{3}} \, .
\end{equation}
Even for a perfectly suitable inflationary potential $V_{\inf}(\phi)$, once the value of $V_{\inf}(\phi)$ becomes large enough, the second term in Eq.~\eqref{Eq:KLProblemVmodPlusVinf} dominates and $\sigma$ becomes a run-away direction. This is actually independent of the particular form of the moduli stabilization sector (for now, it is KKLT), \emph{i.e.} there is always an upper bound for the inflationary energy scale. Empirically, it has been argued that to avoid decompactification for the KKLT moduli stabilization scheme, we have to require \cite{Kallosh:2004yh}
\begin{equation}
  \label{Eq:KLProblemVtotBound}
  V_{\tot} \lesssim \mathcal{O}(1) V_{B} \, .
\end{equation} 

However, for the KKLT scenario, the height of the barrier is related to the gravitino mass in the present vacuum, cf. Eq.~\eqref{Eq:KLProblemRelationGravitinoMassBarrierHeight}, and thus with $\mathcal{H}_{\inf}^{2} \sim V_{\tot}$ the upper bound becomes
\begin{equation}
  \label{Eq:KLProblemHinfBound}
  \mathcal{H}_{\inf} \lesssim m_{3/2}^{\text{today}} \, .
\end{equation}
This is at odds with having a high scale of inflation and low-energy supersymmetry, which requires $\mathcal{H}_{\inf} \gg m_{3/2}^{\text{today}}$ \footnote{In the context of the large volume scenario, the upper bound becomes even more severe, namely $  \mathcal{H}_{\inf} \lesssim \left(m_{3/2}^{\text{today}} \right)^{3/2} $ in Planck units \cite{Conlon:2008cj}.}. 

\begin{figure}[ht]
  \label{Fig:KKLTDestabilization}
  \begin{center}
  \includegraphics[width=0.7\textwidth]{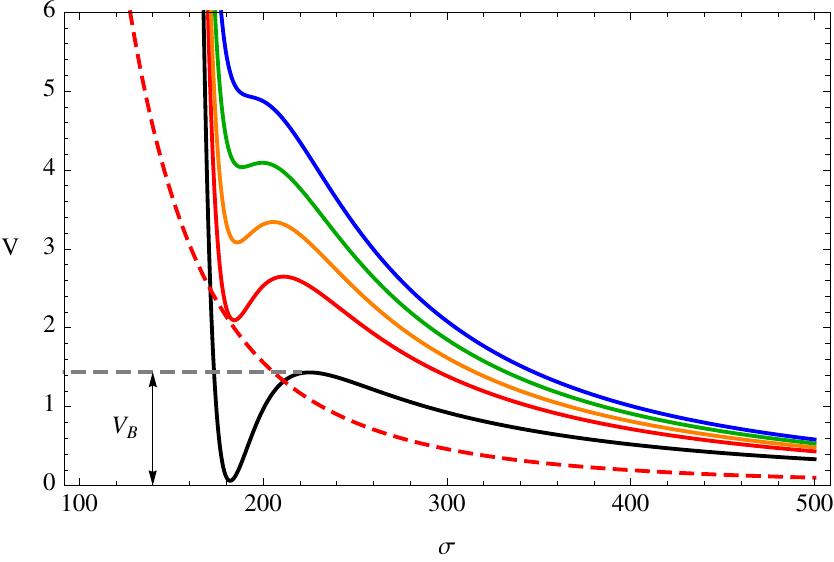}
  \caption{Schematic picture of the destabilization of the uplifted minimum due to adding an inflationary sector. The moduli potential after inflation is given by the solid black line. The dashed red line corresponds to the contribution from only $V_{\inf}(\phi) / \sigma^{3}$, while the solid red line corresponds to $V_{\mod}(\sigma) + \sigma^{-3} \, V_{\inf}(\phi)$. The other solid colored lines correspond to different increasing values of $V_{\inf}(\phi)$. The dashed grey line indicates the height of the barrier $V_{B}$. The scale on the vertical axis is arbitrary.}
  \end{center}
\end{figure}

We illustrate the destabilization of the uplifted minimum for the modulus due to inflation in Fig.~\ref{Fig:KKLTDestabilization}. Obviously, by increasing $V_{\inf}(\phi)$, for a certain critical value of $V_{\inf}(\phi)$ the minimum disappears and the modulus runs away to infinity.

The solution suggested by KL \cite{Kallosh:2004yh} is to use a form of the superpotential for which the gravitino mass in the present vacuum is unrelated to the height of the barrier. To achieve this decoupling, the terms in the superpotential must be fine-tuned.

The crucial issue of the KL problem arises from the notion of having a single, common scale to the moduli stabilization mechanism {\it during} and {\it after} inflation. In this article, we propose a scenario where two \emph{different} mechanisms stabilize the modulus during and after inflation. In particular, we consider models where a certain term in the K\"ahler potential is responsible for stabilizing the volume modulus during inflation, whereas a standard KKLT-type superpotential stabilizes the modulus after inflation, and thereby sets the gravitino mass in the present vacuum. In the next section, we will outline the general setup of our scenario, followed by an explicit example.

\section{Resolution: A General Framework}
\label{Sec:ResolutionGeneralFramework}

We consider generic superpotentials of the following form
\begin{equation}
  \label{Eq:ResolutionGenericSuperpotential}
  W = W_{\inf}(X, \Phi, \dots) + W_{\mod}(T) \, ,
\end{equation}
with $X$ denoting the field whose F-term drives the inflationary vacuum energy and $\Phi$ containing the inflaton. The dots in the argument of $W_{\inf}$ represent possible other fields, e.g. some waterfall fields to realize hybrid inflation. The F-term potential during inflation takes the form
\begin{equation}
  \label{Eq:ResolutionGenericFtermPotential}
  V_{F} = e^{K} \, K^{X \bar{X}} \lvert D_{X} W_{inf} \rvert^{2} + V_{\mod}(\sigma) + V_{\mix}  \, ,
\end{equation}
where $V_{\mod}(\sigma)$ originates solely from the modulus sector $W_{\mod}(T) $ and is responsible for moduli stabilization only {\it after inflation} when the F-term of $X$ (the vacuum energy) - the first term in Eq.~\eqref{Eq:ResolutionGenericFtermPotential} - has vanished. Since the Hubble scale at the end of inflation is much smaller than the Hubble scale during inflation, $W_{\mod}(T) $ need not be necessarily large in contrast to the usual setup where $W_{\mod}(T) $ is responsible for moduli stabilization both during and after inflation. $V_{\mix}$ denotes possible additional mixing terms, in particular, the contributions due to $K_{T \bar{X}} \neq 0$. There are in general also other mixing terms, e.g. due to $K_{\Phi \bar{T}} \neq 0$, but we assume those to be negligible in the following.

During inflation, moduli stabilization is achieved by a suitable moduli-dependence of the first term. Moreover, since we focus on setups where $W_{\inf}$ satisfies 
\cite{Stewart:1994ts,Kawasaki:2000yn,Antusch:2008pn,Antusch:2009ef,Antusch:2009ty,Kallosh:2010ug,Kallosh:2010xz}
\begin{equation}
  \label{Eq:ResolutionWinfConstraints}
  W_{\inf} \approx 0 \,\,\,\, , \,\,\,\, W_{\inf, \, X} \neq 0 \,\,\,\, \text{and} \,\,\,\, W_{\inf, \, i \neq X} \approx 0 \, ,
\end{equation}
the minimum for $T$ during inflation has to be generated by the prefactor of the first term, $e^{K} \, K^{X \bar{X}}$. Below we give a simple example (on which we will be more explicit later in section~\ref{Sec:AnExplicitExample}), where we keep a no-scale K\"ahler potential for $T$ and modify the K\"ahler metric for $X$ in an appropriate way to stabilize the modulus during inflation. But keep in mind that our framework allows for more general possibilities, e.g. if one breaks the no-scale structure in the K\"ahler potential by including $\alpha^{\prime}$-corrections \cite{Becker:2002nn}.

$W_{\mod}(T)$ is responsible for stabilizing the modulus at the end of inflation. Upon including the necessary uplifting term, SUSY is broken in the present vacuum. As long as the Hubble scale after inflation is smaller than the TeV-scale, the modulus will always remain stable since TeV-scale SUSY breaking generically induces moduli masses which are of the same order or heavier. Therefore, we can decouple low-energy SUSY breaking from the inflationary scale, thereby evading the KL problem, that is we can generically assume $\lvert W_{\mod} \rvert, \lvert D_{T} W_{\mod} \rvert \ll \lvert W_{\inf, X} \rvert$.

From now on, we restrict ourselves to a particular form of the K\"ahler potential, which contains the aforementioned coupling between $T$ and $X$, namely
\begin{equation}
  \label{Eq:ResolutionTXKaehlerPotential}
  K = - 3 \ln (T + \bar{T}) + \lvert X \rvert^{2} \left(1 - \beta (T + \bar{T}) - \gamma \lvert X \rvert^{2} \right) + \dots \, .
\end{equation}
It has been argued in \cite{Antusch:2011ei} that couplings qualitatively similar to the second term in the brackets in Eq.~\eqref{Eq:ResolutionTXKaehlerPotential} can arise as moduli-dependent string loop corrections in heterotic orbifold compactifications. For simplicity, we have dropped a possible overall factor of $(T~+~\bar{T})^{-p}$ (with $p$ some rational number) for the terms in the brackets in Eq.~\eqref{Eq:ResolutionTXKaehlerPotential}, which does not change the results significantly. A discussion of the origin of such terms in type IIB string theory is beyond the scope of our present paper and we defer it to the future.
 
Due to the $\lvert X \rvert^{4}$ term, $X$ generically acquires a large mass during inflation \cite{Kawasaki:2000yn} for $\gamma \gtrsim \mathcal{O}(1)$\footnote{There exists a nice geometric interpretation for this requirement in terms of the sectional curvature along the Goldstino direction, cf. \cite{GomezReino:2006dk,GomezReino:2006wv,GomezReino:2007qi,Covi:2008ea,GomezReino:2008px,Covi:2008cn}.}
 and remains near zero\footnote{Note that $X$ is not stabilized exactly at zero since the non-vanishing gravitino mass $m_{3/2}^{2} \propto \lvert W_{\mod} \rvert^{2}$ induces a shift away from zero. However, this shift is parametrically small for small $\lvert W_{\mod} \rvert$, cf. Eq.~\eqref{Eq:LeadingOrderDeltaXImInflationKKLTShiftChaotic}.}. Schematically, for $\lvert X \rvert \ll 1$, $V_{F}$ is given by
\begin{equation}
  \label{Eq:ResolutionSchematicFtermPotential}
  V_{F} \sim \frac{\lvert W_{X} \rvert^{2}}{\sigma^{3} \, (1 - 2 \, \beta \,
\sigma)} + V_{\mod}(\sigma) + \mathcal{O}(X) \, .
\end{equation}
In the limit $W_{\mod} \rightarrow 0$, we have $X = 0$ and the potential during inflation  is entirely given by the first term. Thus, let us concentrate on the first term for a moment: If $\beta > 0$, this term stabilizes $\sigma$ at $\sigma_{\inf} \sim \beta^{-1}$ with a large mass proportional to $\Hinf^{2}~\sim~\lvert W_{X} \rvert^{2}/\sigma_{\inf}^{3}$. More precisely, taking into account the non-canonical kinetic term for $T$, we find $m_{T}^{2}~\simeq~\mathcal{O}(10) \, \Hinf^{2}$. At the end of inflation, the vaccum energy goes away and the modulus would become unstable. However, in this phase, $V_{\mod}(\sigma)$ begins to dominate and stabilizes the modulus. The novel feature of this setup is that the mechanisms for moduli stabilization during and after inflation are a priori unrelated. Therefore, the gravitino mass in the present vacuum is independent of the inflationary scale, which evades the KL problem.

The presence of the moduli sector induces potentially dangerous corrections to the inflationary trajectory. However, for low-energy SUSY breaking (in particular for TeV-scale SUSY breaking) and high-scale inflation, these corrections are parametrically small, as we show in section~\ref{Sec:ExampleCorrectionsInflation} for a KKLT-type superpotential and in appendix~\ref{Sec:AppGeneralWmodCorrectionsInflation} for a generic choice of $W_{\mod}(T)$.

The price we have to pay is that the minima during and after inflation do not have to be the same. Since there seems to be no dynamical mechanism involved, we choose them to (almost) coincide by tuning the parameters appropriately. This imposes a relation between the parameter $\beta$ and the parameters controlling $W_{\mod}$, but does not affect the possiblity of having low-energy SUSY breaking and high-scale inflation at the same time.

\section{Resolution: An Explicit Example}
\label{Sec:AnExplicitExample}

\begin{figure}[ht]
  \label{Fig:ModuliPotentials}
  \begin{center}
  \includegraphics[width=0.8\textwidth]{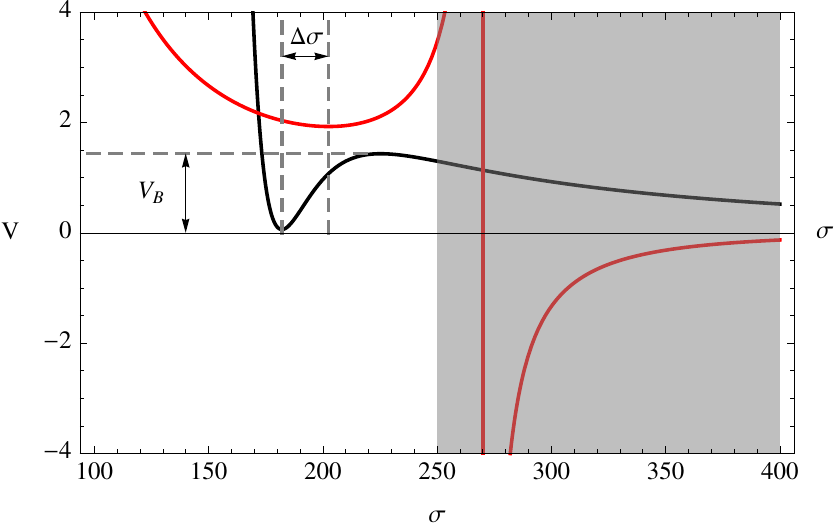}
  \caption{Schematic plot of the two moduli stabilization potentials during inflation (red) and after inflation (black). The grey dashed lines indicate the displacement $\Delta \sigma$ of the two minima and the height of the barrier of the minimum after inflation. The grey region in the right part marks the regime where our effective field theory at second order in the derivatives ceases to be valid. The scale on the vertical axis is arbitrary.}
  \end{center}
\end{figure}

In this section, we illustrate our general idea in a simple toy model: shift symmetric chaotic inflation combined with a KKLT-type superpotential. Results for hybrid (or tribrid) models and inflationary scenarios based on a Heisenberg symmetry will appear elsewhere \cite{HeisenbergTribrid}.


For simplicity, we consider a chaotic inflation model \cite{Kawasaki:2000yn} based on the superpotential
\begin{equation}
  \label{Eq:WInfChaotic}
  W_{\inf}(\Phi, X) = m \, X \, \Phi \, ,
\end{equation}
where $\Phi$ contains the inflaton and $X$ is the field whose F-term drives the inflationary vacuum energy. For the modulus sector, we consider a KKLT-type superpotential \cite{Kachru:2003aw},
\begin{equation}
  \label{Eq:WmodKKLT}
  W_{\mod}(T) = w_{0} + A e^{- a \, T} \, .
\end{equation}
We consider a no-scale K\"ahler potential for $T$ and the coupling between $T$ and $X$ introduced in Eq.~\eqref{Eq:ResolutionTXKaehlerPotential}. To solve the supergravity $\eta$-problem, we assume a shift symmetry for the inflaton direction.\footnote{We do not discuss the naturalness of the shift symmetry with respect to quantum (gravity) corrections here. For our purposes, we assume a solution to the $\eta$-problem and effectively parametrize the resulting inflaton potential by $W_{\inf}$ in Eq.~\eqref{Eq:WInfChaotic} and a negligible breaking of the shift symmetry in the K\"ahler potential Eq.~\eqref{Eq:KModInfShiftChaotic}.} That is, we consider the K\"ahler potential
\begin{equation}
  \label{Eq:KModInfShiftChaotic}
  K = - 3 \ln \left( T + \bar{T} \right) + \frac{1}{2} \left(\Phi + \bar{\Phi} \right)^{2} + \lvert X  \rvert^{2} - \beta \left( T + \bar{T} \right) \lvert X \rvert^{2} - \gamma \lvert X \rvert^{4} \, .
\end{equation}
The first term is invariant under a shift symmetry for the imaginary part of $\Phi$, which protects this direction from the supergravity $\eta$-problem. Recall that the last term ensures that $X$ is stabilized near $X = 0$ with $m_{X} \gtrsim \Hinf$ during inflation if $\gamma \gtrsim \mathcal{O}(1)$. The coupling between $T$ and $X$ in Eq.~\eqref{Eq:KModInfShiftChaotic} stabilizes $T$ during inflation, while the superpotential $W_{\mod}$ fixes $T$ after inflation. As noted in section~\ref{Sec:ReviewKalloshLindeProblem}, since the minimum generated by $W_{\mod}$ is a supersymmetric AdS minimum, we need to uplift it to a dS minimum with a tiny cosmological constant. To achieve this, an uplifting contribution is added to the potential,
\begin{equation}
  \label{Eq:Vuplift}
  V_{\up} = \frac{c_{\up}}{\left( T + \bar{T} \right)^{2}} \, ,
\end{equation}
which is motivated by introducing anti-D$3$-branes at the tip of a warped throat in a string theory realization of such a setup. Here, however, we do not refer to a particular string theory embedding of our scenario and simply view the above setup as an effective parametrization for the potential of the modulus $T$ and the inflationary sector with $X$ and $\Phi$.

Note that we assume the absence of any mixing between $T$ and $\Phi$ and only a mixing between $T$ and $X$ given by the second term in the K\"ahler potential in Eq.~\eqref{Eq:KModInfShiftChaotic}. We comment on possible effects of such a mixing later on.

We denote the real and imaginary parts of the scalar components of the chiral superfields as follows
\begin{equation}
  \label{Eq:RealFldsShiftChaotic}
  T \equiv \sigma + i \, \alpha \, , \, \Phi \equiv \phi_{R} + i \, \phi_{I} \, , \, X
\equiv x_{R} + i \, x_{I} \, .
\end{equation}
The vacuum expectation values of $\sigma$ during and after inflation are denoted by $\sigma_{\inf}$ and $\sigma_{0}$, respectively. In addition, we choose the phases of the parameters in $W_{\mod}$ such that the minimum is at $\alpha = 0$. More precisely, for the superpotential in Eq.~\eqref{Eq:WmodKKLT}, we choose $w_{0}$ to be real and negative and $A$ to be real and positive.

A schematic plot of the two moduli stabilizing potentials generated by the F-term of $X$ (cf. the first term in Eq.~\eqref{Eq:ResolutionSchematicFtermPotential}) and the one induced by the KKLT-type superpotential and the uplifting term (cf. Eqs.~\eqref{Eq:WmodKKLT} and \eqref{Eq:Vuplift}) can be found in Fig.~\ref{Fig:ModuliPotentials}.

To analyze the effects of the presence of a general $W_{\mod}$ during inflation, the basic idea is to perform a perturbative expanson in $W_{\mod}$, $D_{T} W_{\mod}$, $W_{\mod}^{\prime \prime}$ etc., with dimensionless expansion parameters given by $W_{\mod} / W_{X}$ etc. (up to appropriate powers of $M_{P}$). They parametrize the impact of the modulus sector on the inflationary trajectory and for low-energy SUSY breaking and high-scale inflation these expansion parameters are parametrically small, thereby making our treatment self-consistent. For the KKLT-type superpotential in Eq.~\eqref{Eq:WmodKKLT}, this procedure is essentially equivalent to an expansion in small $\lvert w_{0} \rvert \ll 1$ and large $a \, \sigma \gg 1$. Results for a general $W_{\mod}(T)$ are presented in appendix \ref{Sec:AppendixGeneralWmod}.

\subsection{Stability of the Vacuum After Inflation}
\label{Sec:ExampleVacuumStability}

We now discuss two possible problems which lead to constraints on the parameter space. Firstly, since the two minima during and after inflation generically do not coincide, we may not end up in the true minimum. In particular, we may overshoot the minimum and run away to infinity. Secondly, without $W_{\mod}$ and $V_{\up}$, the vacuum after inflation is $\Phi = X = 0$ and there are two flat directions corresponding to the real and imaginary parts of $T$. If we add $V_{\mod}$, these two flat directions are stabilized. However, we may induce some instabilities for $\Phi$ and $X$. The implications of these two issues for the model parameters are discussed in sections~\ref{Sec:ExampleOvershoot} and \ref{Sec:ExampleStabilityBound}, respectively.

\subsubsection{Overshoot Problem}
\label{Sec:ExampleOvershoot}

To avoid overshoot of the modulus after inflation (as long as there is no dynamical mechanism implemented for a smooth transition between the two minima) we will require that the positions of the minima are sufficiently close to each other, i.e. $\sigma_{0} \approx \sigma_{\inf}$ (c.f.\ Fig.~\ref{Fig:ModuliPotentials}). This relates the parameters controlling $W_{\mod}$ to the parameter $\beta$ in the K\"ahler potential. For the specific case of the KKLT superpotential, Eq.~\eqref{Eq:WmodKKLT}, if we ignore the shift due to the uplifting potential Eq.~\eqref{Eq:Vuplift}, we can estimate $\sigma_{0}$ as\footnote{Actually, there is a slighlty better approximation for the position of the AdS-minimum, namely
\begin{equation}
  \nonumber
  \sigma_{0} \simeq - \frac{1}{a} \ln \left\lvert \frac{w_0}{A} \frac{1}{1-\frac{2}{3}
\ln \lvert \frac{w_0}{A} \rvert} \right\rvert \, .
\end{equation}
However, for our purposes here, the approximation in Eq.~\eqref{Eq:KKLTMinimumApprox} is sufficient since we neglect the uplifting and only need a rough estimate for $\sigma_{0}$.
}
\begin{equation}
  \label{Eq:KKLTMinimumApprox}
  \sigma_0 \simeq -\mathcal{O}(1) \frac{1}{a} \ln
\left\lvert \frac{w_{0}}{A} \right\rvert \, .
\end{equation}
During inflation, $V_{\mod}$ induces only a tiny shift which we can neglect and the position of the minimum is approximately given by
\begin{equation}
  \label{Eq:InfMinimumApprox}
  \sigma \simeq \sigma_{\inf} \equiv \frac{3}{8 \beta} \, .
\end{equation}
For the KKLT superpotential, one typically uses $A \simeq 1$
and $a \simeq \frac{2 \pi}{N}$ for some integer\footnote{For the effective field theory to be valid, we must have $N \gg 1$ such that $\sigma \gg 1$.} $N$ such that effectively mainly $w_{0}$ determines the value of $\sigma_{0}$ and thus also the size of the gravitino mass after inflation.

For the two minima to be roughly at the same position, $\sigma_{0} \approx \sigma_{\inf}$, we have to tune the parameters such that
\begin{equation}
  \label{Eq:w0betaRelation}
  \frac{1}{\beta} \simeq - \mathcal{O}(1) \frac{1}{a} \ln \left\lvert \frac{w_{0}}{A}
\right\rvert \, .
\end{equation}
We will assume this condition to be satisfied to a good approximation.

\subsubsection{Stability Bound on $w_0$}
\label{Sec:ExampleStabilityBound}

There is yet another condition leading to a bound on $w_0$, as we now discuss. Let us start by noting that, 
due to the absence of any mixings, the masses for $\phi_{R}$ and $\phi_{I}$ are simply given by
\begin{equation}
  \label{Eq:PhiRMassKKLTShiftChaotic}
  \frac{\partial^{2} V}{\partial \phi_{R}^{\, 2}} = \frac{m^{2}}{8 \, 
\sigma_{0}^{\, 3} \, (1 - 2 \, \beta \, \sigma_{0})} - \frac{\lvert W_{\mod}(\sigma_{0}) \rvert^{2}}{4 \, \sigma_{0}^{\, 3}} \, ,
\end{equation}
and
\begin{equation}
  \label{Eq:PhiIMassKKLTShiftChaotic}
  \frac{\partial^{2} V}{\partial \phi_{I}^{\, 2}} = \frac{m^{2}}{8 \, 
\sigma_{0}^{3} \, (1 - 2 \, \beta \, \sigma_{0})} \, ,
\end{equation}
respectively. To avoid a tachyonic mass for $\phi_{I}$, we have to require that $\sigma_{0}$ satisfies
\begin{equation}
  \label{Eq:T0ValueBoundKKLTShiftChaotic}
  \sigma_{0} < \frac{1}{2 \beta} \, .
\end{equation}
Note that this condition is also necessary to avoid a wrong sign for the kinetic term of $X$\footnote{For values too close to the upper bound, i.e. $\sigma_{0} \approx \frac{1}{2 \beta}$, it is no longer justified to work only at second order in the derivatives and our effective field theory ceases to be valid.}, independently of the stability condition for $\phi_{I}$. From also avoiding an instability for $\phi_{R}$, we get an upper bound on the size of $W_{\mod}$ at the minimum in terms of $m$:
\begin{equation}
  \label{Eq:WmodBoundVacStabShiftChaotic}
  \lvert W_{\mod}(\sigma_{0}) \rvert^{2} \lesssim \frac{m^{2}}{2 \, (1 - 2 \, \beta \, \sigma_{0})} \approx 2 \, m^{2} \, ,
\end{equation}
where the last step assumes $\sigma_{0} \approx \sigma_{\inf} \equiv \frac{3}{8 \beta}$. There is no further constraint from the stability of $x_{R,I}$. Using the superpotential in Eq.~\eqref{Eq:WmodKKLT}, the bound in Eq.~\eqref{Eq:WmodBoundVacStabShiftChaotic} becomes a bound on $w_{0}$ in terms of $m$:
\begin{equation}
  \label{Eq:WmodBoundVacStabKKLTShiftChaotic}
  w_{0}^{\, 2} \lesssim \frac{(3 + 2 \, a \, \sigma_{0})^{2}}{8 \, a^{2} \, \sigma_{0}^{\, 2} \, (1 - 2 \, \beta \, \sigma_{0})} \, m^{2} \approx \, 2 \, m^{2} + \mathcal{O}\left( \frac{m^{2}}{a \, \sigma_{0}} \right) \, ,
\end{equation}
where in the last step we again assumed $\sigma_{0} \approx \sigma_{\inf}$ and expanded for $a \, \sigma_{0} \gg 1$. Note that this bound will not be affected by adding the uplifting sector as long as the uplifting term $V_{\up}$, Eq.~\eqref{Eq:Vuplift}, does not depend on $\Phi$. Moreover, if any mixing between $T$ and $\Phi$ would be present, the bound is not directly on $W_{\mod}$, but on a particular combination of $W_{\mod}$ and $D_{T} W_{\mod}$ depending on the mixing terms. For the KKLT superpotential and the no-scale K\"ahler potential, one typically has $\lvert D_{T} W_{\mod}(\sigma_{0}) \rvert \sim \lvert W_{\mod}(\sigma_{0}) \rvert$ at the uplifted minimum and thus the bound would be essentially again on the size of $\lvert W_{\mod}(\sigma_{0}) \rvert$.

The important lesson here is that vacuum stability puts some constraint on the size of supersymmetry breaking after inflation, but for high-scale inflation still allows for a large range of possible values of $w_{0}$, in particular those leading to low-energy SUSY breaking.

\subsection{Comment on the Cosmological Moduli Problem}
\label{Sec:ExampleModuliProblem}

Since the moduli stabilization mechanisms during and after inflation are of completely
different origin, the minima for the modulus field
during and after inflation generically do not exactly coincide. This means the modulus field will oscillate and eventually dominate the universe. If it decays too late in the history of the universe, in particular if it decays after Big Bang Nucleosynthesis (BBN), it causes the well-known cosmological moduli problem \cite{Coughlan:1983ci}.  However, since for the KKLT case there is a little hierarchy of scales \cite{Choi:2004sx,Choi:2005ge,Choi:2005uz,Choi:2006im,Abe:2005rx,Abe:2007yb},
\begin{equation}
  m_{T} \sim 16 \pi^{2} \, m_{3/2} \sim (16 \pi^{2})^{2} \, m_{soft} \, ,
\end{equation}
we can assume that the modulus (and also the gravitino) is heavier than about $30$ TeV such that it will decay before BBN and thus the cosmological moduli problem is avoided.  We leave a more detailed investigation of both the non-thermal history and the issue of relaxing the tuning $\sigma_{0} \approx \sigma_{\inf}$, e.g. by implementing a dynamical mechanism, for the future.

\subsection{Corrections to the Inflationary Trajectory}
\label{Sec:ExampleCorrectionsInflation}

The inflationary trajectory is shifted due to the presence of the modulus sector. If we neglect $W_{\mod}$ and $V_{\up}$, the trajectory is given by $\sigma = \sigma_{\inf} \equiv \frac{3}{8 \beta}$, $\alpha = 0$, \footnote{\label{Foot:FrozenAxion} Without $W_{\mod}$, $\alpha$ is not fixed at zero, but effectively frozen since it becomes massless in the limit $W_{\mod} \rightarrow 0$. As noted above, we choose the phases in $W_{\mod}$ such that it has a minimum at $\alpha =  0$. Thus, once we include $W_{\mod}$, the minimum both during and after inflation is at $\alpha = 0$.} $X = 0$, $\phi_{R} = 0$ and $\phi_{I} \neq 0$. Except for the derivative with respect to $\phi_{I}$, the only other non-vanishing derivatives along this trajectory are
\begin{equation}
  \label{Eq:TShiftInflationTrajectoryKKLTShiftChaotic}
  \frac{\partial V}{\partial \sigma} = - \frac{A^{2} \, a \, e^{-2 \, a \, \sigma} \left(6 + 7 \, a \, \sigma +2 \, a^{2} \, \sigma^{2} \right)}{6 \, \sigma^{3}} - \frac{w_{0} \, A \, c \, e^{- a \, \sigma} \left(2 + a \, \sigma \right)}{2 \, \sigma^{3}} - \frac{3 \, a^{2} \, \sigma_{0} \, w_{0}^{2}}{\sigma^{3} \, \left(3 + 2 \, a \, \sigma_{0} \right)^{2}} \, ,
\end{equation}
and
\begin{equation}
  \frac{\partial V}{\partial x_{I}} = \frac{m \, \phi_{I}}{4 \, \sigma^{3}} \left(w_{0} + A e^{- a \, \sigma} \right) \, .
\end{equation}
To (slighlty over-) compensate the negative cosmological constant in the would-be AdS-vacuum, we tune the uplifting potential $V_{\up}$ by choosing (recall that $\sigma_{0} \approx \sigma_{\AdS}$)
\begin{equation}
  c_{\up} \simeq \frac{3 \lvert W_{\mod}(\sigma_{0}) \rvert^2}{2 \, \sigma_{0}} = \frac{6 \, a^{2} \, \sigma_{0} \, w_{0}^{2}}{(3 + 2 \, a \, \sigma_{0})^{2}} \, .
\end{equation}
Hence, we see that the last term in Eq.~\eqref{Eq:TShiftInflationTrajectoryKKLTShiftChaotic} is precisely the contribution from $V_{\up}$.

As mentioned above, we compute the shifts in a perturbative expansion for $\lvert w_{0} \rvert \ll 1$ and $a \, \sigma \gg 1$. At leading order, the shifts in $\sigma$ and $x_{I}$ during inflation are given by
\begin{equation}
  \label{Eq:LeadingOrderDeltaTInflationKKLTShiftChaotic}
  \begin{split}
  \delta \sigma \simeq \,\, & \frac{\sigma_{\inf} \, w_{0}^{2}}{2 \, m^{2} \, (1 + 32 \, \gamma \, \phi_{I}^{2})} \\
  & + \frac{A^{2} \, e^{-2 \, a \, \sigma_{\inf}} (3 \phi_{I}^{2} + a \, \sigma_{\inf} \, (2 + (3 + 64 \gamma) \phi_{I}^{2}))}{6 \, m^{2} \, \phi_{I}^{2} \, (1 + 32 \, \gamma \, \phi_{I}^{2})} \\
  & + \frac{A \, e^{-a \, \sigma_{\inf}} \, \sigma_{\inf} \, w_{0} \, (6 \phi_{I}^{2} + a \, \sigma_{\inf} \, (2 + (3 + 64 \gamma) \phi_{I}^{2}))}{6 \, m^{2} \, \phi_{I}^{2} \, (1 + 32 \, \gamma \, \phi_{I}^{2})} \, ,
  \end{split}
\end{equation}
and
\begin{equation}
  \label{Eq:LeadingOrderDeltaXImInflationKKLTShiftChaotic}
  \delta x_{I} \simeq  -\frac{\phi_{I} \left( W_{\mod} + \overline{W}_{\mod}
\right)}{m \, (1 + 32 \, \gamma \, \phi_{I}^{2})}  \simeq - \frac{2 \, \phi_{I} \, (w_{0} + A e^{- a \, \sigma_{\inf}})}{m \, (1 + 32 \, \gamma \, \phi_{I}^{2})} \, .
\end{equation}
Note that the shift $\delta x_{I}$ vanishes as $\phi_{I} \rightarrow 0$ such that $\Phi = X = 0$ at the end of inflation. More interestingly, this shift is entirely controlled by the value of the gravitino mass during inflation since for $\phi_{I} \gg 1$
\begin{equation}
  \label{Eq:LeadingOrderDeltaXImInflationKKLTShiftChaoticGravitinoMass}
  \delta x_{I} \simeq - \frac{\phi_{I} \, (W_{\mod} + \overline{W}_{\mod})}{m \, (1 + 32 \, \gamma \, \phi_{I}^{2})} \sim - \frac{m_{3/2} \, \sigma_{\inf}^{3/2}}{m \, \phi_{I}} \, ,
\end{equation}
with $m_{3/2}$ given by the usual expression
\begin{equation}
  m_{3/2}^{2} = e^{\langle K \rangle} \, \lvert \langle W \rangle \rvert^{2} \, .
\end{equation}
Due to the suppression by the large inflationary F-term $W_{X} \simeq m \phi_{I}$, $\delta x_{I}$ is parametrically small for small values of the gravitino mass.

Including these shifts, the potential along the inflationary trajectory is given by
\begin{equation}
  \label{Eq:InflatonPotentialKKLTShiftChaotic}
  V_{\inf} \simeq \frac{m^{2} \, \phi_{I}^{2}}{4 \sigma_{\inf}^{\, 3}} + V_{\mod}(\sigma_{\inf}) + \frac{\phi_{I}^{2} \, (w_{0} + A e^{- a \, \sigma_{\inf}})^{2}}{4 \, \sigma_{\inf}^{\, 3} \, (1 + 32 \, \gamma \, \phi_{I}^{2}))} \, ,
\end{equation}
where the first term is the standard contribution (up to the factor of $\sigma_{\inf}^{3}$) and the last term corresponds to the contribution from $\sim m^{2} \, x_{I}^{2}$ since now $x_{I} \neq 0$. The other terms combine to the moduli potential $V_{\mod}$, which is induced by $W_{\mod}$ and $V_{\up}$, evaluated at $\sigma_{\inf}$. Recall that to ensure $m_{X} \gtrsim \mathcal{H}$ during inflation we require $\gamma \gtrsim \mathcal{O}(1)$. Thus, the inflaton-dependence of the potential at large values of $\phi_{I}$ is not significantly affected by the addition of the last term in Eq.~\eqref{Eq:InflatonPotentialKKLTShiftChaotic}. Together with the bound from Eq.~\eqref{Eq:WmodBoundVacStabKKLTShiftChaotic}, $w_{0}^{\,2} \lesssim 2 \, m^{2}$, and the assumption $\sigma_{0} \approx \sigma_{\inf}$, we can already  anticipate at this stage that no large corrections are expected.

The next step is to compute the corrections to the inflationary observables. Note that the parameter $m$ has to be redefined, both due to the factor of $\sigma_{\inf}^{3}$ in the first term of Eq.~\eqref{Eq:InflatonPotentialKKLTShiftChaotic} and due to the other additional terms. As usual, it is fixed by matching the amplitude of the scalar perturbations to the observed value.

Assuming inflation ends at $\phi_{I} \approx 0$, we can compute the number of
e-folds $N_{e}$ in terms of $\phi_{I}$. The two slow-roll parameters which determine the
inflationary predictions are
\begin{equation}
  \label{Eq:InflationEpsilonEtaDefinition}
  \epsilon = \frac{1}{2} M_{P}^{2} \left( \frac{V^{\prime}}{V} \right)^{2} \,\,\,\, \text{and} \,\,\,\, \eta = M_{P}^{2} \frac{V^{\prime \prime}}{V} \, .
\end{equation}
The amplitude of the scalar power spectrum is given by 
\begin{equation}
  \mathcal{P}_{\mathcal{R}}^{1/2}  = \frac{1}{2 \sqrt{3 \pi}} \frac{V^{3/2}}{\lvert V'\rvert}
\end{equation}
As usual, the parameter $m$ is fixed by matching the observed value for the amplitude of the scalar power spectrum to the predicted one. To leading order, the amplitude of the scalar power spectrum $\mathcal{P}_{\mathcal{R}}^{1/2}$ is given by
\begin{equation}
  \label{Eq:InflationScalarAmplitudeNeFoldsKKLTShiftChaotic}
  \begin{split}
  \mathcal{P}_{\mathcal{R}}^{1/2} \simeq \, & \frac{m \, N_{e}}{2 \, \sqrt{3} \, \pi \, \sigma_{\inf}^{3/2}} \Bigg( 1 + \frac{w_{0}^{2}}{m^{2} \, N_{e}} \left( \frac{\sigma_{\inf}}{\sigma_{0}} \left(\frac{9}{16} - \frac{3}{8} \ln(4 N_{e})\right) + \frac{2 \ln(128 \, \gamma \, N_{e}) - 3}{256 \, \gamma} \right) \\ & + \frac{w_{0} A e^{-a \, \sigma_{\inf}}}{m^{2} N_{e}} \left( - \frac{a \, \sigma_{\inf}}{4} \left(2 \ln(4 N_{e}) - 3\right) + \frac{2 \ln(128 \, \gamma \, N_{e}) - 3}{128 \, \gamma} \right) \\ & + \frac{A^{2} e^{-2  \, a \, \sigma_{\inf}}}{m^{2} \, N_{e}} \bigg( - \frac{a^{2} \, \sigma_{\inf}^{2}}{12} \left(2 \ln(4 N_{e}) - 3\right) - \frac{a \, \sigma_{\inf}}{4} \left(2 \ln(4 N_{e}) - 3 \right) \\ & + \frac{2 \ln(128 \, \gamma \, N_{e}) - 3}{256 \, \gamma} \bigg) \Bigg)\, .
  \end{split}
\end{equation}
Obviously, compared to the inflationary scenario without the modulus sector, we only need to redefine $m$ to account for the prefactor $\sigma_{\inf}^{3/2}$. The extra terms are all negligible and only would give rise to higher order corrections in the expressions for $\epsilon$ and $\eta$ so we can ignore those.

Having fixed the parameter $m$, we can calculate the predictions for the observables. For example, the scalar spectral index $n_s = 1 - 6 \, \epsilon + 2 \, \eta$ is given by
\begin{equation}
  \label{Eq:InflationScalarIndexNeFoldsKKLTShiftChaotic}
  \begin{split}
  n_{s} - 1 \simeq \, & - \frac{2}{N_{e}} + \frac{w_{0}}{m^{2} \, N_{e}^{2}} \left(- \frac{3 \, \sigma_{\inf} \, \left( 2 \ln(4 N_{e}) - 5 \right)}{8 \, \sigma_{0}} + \frac{2 \ln(128 \, \gamma \, N_{e})}{128 \, \gamma} \right) \\ & + \frac{w_{0} A e^{- a \, \sigma_{inf}}}{m^{2} \, N_{e}^{2}} \left( -\frac{a \, \sigma_{\inf}}{2} (2 \ln(4 N_{e}) - 5) + \frac{2 \ln(128 \, \gamma \, N_{e})}{64 \, \gamma} \right) \\ & \frac{A^{2} e^{- 2 \, a \, \sigma_{\inf}}}{m^{2} \, N_{e}^{2}} \bigg( \frac{a^{2} \, \sigma_{\inf}^{2}}{6} \left(2 \ln(4 N_{e}) - 5 \right) - \frac{a \, \sigma_{\inf}}{2} \left(2 \ln(4 N_{e}) - 5 \right) \\ & + \frac{2 \ln(128 \, \gamma \, N_{e})}{128 \, \gamma}  \bigg) \, ,
  \end{split}
\end{equation}
and the tensor-to-scalar ratio $r = 16 \, \epsilon$ is given by
\begin{equation}
  \label{Eq:InflationTensorToScalarRatioNeFoldsKKLTShiftChaotic}
  \begin{split}
  r \simeq \, & \frac{8}{N_{e}} + \frac{w_{0}^{2}}{m^{2} \, N_{e}^{2}} \left( \frac{3 \, \sigma_{\inf} (\ln(4 N_{e}) - 2)}{\sigma_{0}} - \frac{\ln(128 \, \gamma \, N_{e}) - 2}{16 \, \gamma} \right) \\ & + \frac{w_{0} A e^{- a \, \sigma_{\inf}}}{m^{2} \, N_{e}^{2}} \left( 4 \, a \, \sigma_{\inf} (\ln(4 N_{e}) - 2) - \frac{\ln(128 \, \gamma \, N_{e}) - 2}{8 \, \gamma} \right) \\ & + \frac{A^{2} e^{- 2 a \, \sigma_{\inf}}}{m^{2} \, N_{e}^{2}} \bigg( \frac{4 \, a^{2} \, \sigma_{\inf}^{2}}{3} (\ln(4 N_{e}) - 2) + 4 \, a \, \sigma_{\inf} (\ln( 4 N_{e} ) - 2) \\ & - \frac{\ln(128 \, \gamma \, N_{e}) - 2}{16 \, \gamma} \bigg) \, .
  \end{split}
\end{equation}
Note that all the correction terms start at order $N_{e}^{-2}$ (up to some logarithms). Thus, they are suppressed with respect to the leading contribution $\sim N_{e}^{-1}$, as expected since we perform a perturbative expansion in $W_{\mod} / W_{X}$ etc. and $W_{X}^{2} = m^{2} \, \phi_{I}^{2} \sim m^{2} \, N_{e}$. This suppression is sufficient to keep the corrections induced by the modulus sector small even if we saturate the bound in Eq.~\eqref{Eq:WmodBoundVacStabKKLTShiftChaotic}. Most importantly, for high-scale inflation and low-energy supersymmetry, the inflationary predictions are not significantly affected: the corrections are suppressed by $m_{3/2}^{2} / F_{X}^{2}$ and $F_{T}^{2}/F_{X}^{2}$.

\section{Conclusions}
\label{Sec:Conclusion}

In this article, we have proposed a general scenario for moduli stabilization where low-energy supersymmetry breaking can be 
accommodated together with a high scale of inflation. 
In our proposal the KL problem (which is reviewed in section \ref{Sec:ReviewKalloshLindeProblem}) is resolved because the stabilization of the modulus field \emph{during} and \emph{after} inflation is not associated with a single, common scale, but instead relies on two different mechanism to stabilize the modulus during and after inflation. 

More explicitly (c.f.\ section \ref{Sec:ResolutionGeneralFramework}), we suggest to consider a K\"ahler potential which features a coupling between the modulus field and the field 
whose F-term drives inflation in such a way that the term $V_{\inf} \sim e^{K} \, K^{X \bar{X}} \lvert D_{X} W \rvert^{2}$ creates a minimum for the modulus which stablizes it with large mass during inflation. After inflation, when $D_{X} W$ vanishes, a ``standard'' mechanism involving nonperturbative terms in the superpotential can take over to stabilize the modulus as usual. The way we avoid the KL problem in this setup works essentially as follows: The gravitino mass $m_{3/2}$ now only sets the scale for moduli stabilization after inflation and therefore remains small all the time. During inflation, the scale for moduli stabilization is set by the inflationary energy scale $\Hinf$ itself and no longer also by $m_{3/2}^{\text{today}}$. This 
allows to consistently combine low scale SUSY and high scale inflation.

There is of course a price to pay: Since the two minima for the modulus during and after inflation generically do not coincide,
we have to make sure that there is no overshoot problem after inflation. This requires, for instance, that the difference between the two minima for the modulus lie not too far apart (c.f.\ section \ref{Sec:ExampleOvershoot}). Without a dynamical mechanism to guarantee a smooth transition between the two minima, achieving $\sigma_{0} \approx \sigma_{\inf}$ may require some amount of tuning of the model parameters. However, notice that also the KL solution requires some tuning to disentangle the height of the barrier from the gravitino mass today. Also note that, for a KKLT-type stabilisation mechanism after inflation, the mass of the modulus amounts about $m_T \sim 16\pi^2 \, m_{3/2}$, which means it can be haevy enough to decay before BBN, avoiding the standard cosmological moduli problem.

We have illustrated our general strategy in a simple model of chaotic inflation with a shift symmetry supplemented by a KKLT-type superpotential and uplifting term (c.f.\ section \ref{Sec:AnExplicitExample}). Moduli stabilization during inflation is achieved considering a rather general K\"ahler potential coupling between the field which provides the inflationary vacuum energy by its F-term and the modulus. We also showed that in the limit of high-scale inflation and low-energy supersymmetry breaking, i.e.~for $W, D_{T} W \ll W_{X}$, the corrections to the inflationary observables from the modulus sector become negligible. The appendix contains results for a general moduli stabilizing superpotential $W_{\mod}(T)$ combined with the simple chaotic inflation model. Results for hybrid (or tribrid) inflation models and models based on a Heisenberg symmetry instead of a shift symmetry will appear elsewhere \cite{HeisenbergTribrid}.

Finally, emphasize that our general strategy may work for more general scenarios. Even though many ingredients of our scenario are motivated from a string theory point of view, we did not consider a particular embedding in a string theory compactification here and defer this discussion to the future.

\subsection*{Acknowledgements}
We thank P.~M.~Kostka for collaboration on the initial stage of this project. We would like to thank Alexander Westphal for several illuminating comments and discussions. S.H.~is supported by the German Science Foundation (DFG) Cluster of Excellence ``Origin and Structure of the Universe''. K.D.~is supported by the German Science Foundation (DFG) within the Collaborative Research Center 676 ``Particles, Strings and the Early Universe''. S.A.~acknowledges support by the Swiss National Science Foundation. 

\section*{Appendix}
\appendix

\section{Some Results for Generic $W_{\mod}(T)$}
\label{Sec:AppendixGeneralWmod}

In this appendix, we present some results for a general $W_{\mod}(T)$. The moduli superpotentials we have in mind are of the form
\begin{equation}
  \label{Eq:AppGeneralWmod}
  W_{\mod}(T) = w_{0} + \sum_{n} A_{n} \, e^{- a_{n} T} \, ,
\end{equation}
but their precise form is irrelevant for our discussion. We still restrict ourselves to the chaotic inflation model from section~\ref{Sec:AnExplicitExample}, i.e. we consider 
\begin{equation}
  \label{Eq:AppGeneralW}
  W = m \, X \, \Phi + W_{\mod}(T) \, ,
\end{equation}
and
\begin{equation}
  \label{Eq:AppKNoTPhiMixing}
  K = - 3 \ln \left( T + \bar{T} \right) + \frac{1}{2} \left(\Phi + \bar{\Phi} \right)^{2} + \lvert X  \rvert^{2} - \beta \left( T + \bar{T} \right) \lvert X \rvert^{2} - \gamma \lvert X \rvert^{4} \, .
\end{equation}
If necessary, we allow for the possibility of adding an uplifting term of the form
\begin{equation}
  \label{Eq:AppVuplift}
  V_{\up} = \frac{c_{\up}}{\left( T + \bar{T} \right)^{2}} \, ,
\end{equation}
where the constant $c_{\up}$ is tuned to have an (almost) vanishing cosmological constant. Here, we do not consider other possibilites for uplifting such as those discussed e.g. in \cite{Burgess:2003ic,Achucarro:2006zf,Lebedev:2006qq,Dudas:2006gr,Abe:2006xp,Abe:2007yb}. As before, we choose the phases of the parameters in $W_{\mod}$ such that the minimum for the imaginary part of $T$ is at $\alpha = 0$.

The general strategy is to perform a perturbative expansion in $W_{\mod}$, $D_{T} W_{\mod}$, $W_{\mod}^{\prime \prime}$ and higher derivatives to determine the effect of the modulus sector during inflation. Recall that this is an expansion with dimensionless expansion parameters given by $W_{\mod} / W_{X}$ etc. (up to appropriate powers of $M_{P}$), which parametrize the impact on the inflationary trajectory by adding the modulus sector. Note that this expansion breaks down towards the end of inflation since $W_{X}$ vanishes as $\Phi \rightarrow 0$. For simplicity, we assume the absence of any mixing between $T$ and $\Phi$ and only a mixing betweeen $T$ and $X$ given by the second term in Eq.~\eqref{Eq:AppKNoTPhiMixing}.

For many schemes for moduli stabilization, at the minimum $\sigma_{0}$ one has an upper bound $\lvert D_{T} W_{\mod}(\sigma_{0}) \rvert \lesssim \lvert W_{\mod}(\sigma_{0}) \rvert$, e.g. for the KKLT mechanism one has $\lvert D_{T} W_{\mod}(\sigma_{0}) \rvert \sim \lvert W_{\mod}(\sigma_{0}) \rvert$, while for the KL scenario one has $\lvert D_{T} W_{\mod}(\sigma_{0}) \rvert \ll \lvert W_{\mod}(\sigma_{0}) \rvert$. We restrict our discussion to moduli stabilization scenarios which obey such an upper bound.

\subsection{Stability of the Vacuum After Inflation}
\label{Sec:AppGeneralWmodVacuumStability}

If $W_{\mod}$ and $V_{\up}$ are not present, the vacuum after inflation is given by $\Phi = X = 0$ and both the real and the imaginary part of $T$ remain as flat directions. These two flat directions are stabilized upon adding $W_{\mod}$. However, we may induce some instabilities for $\Phi$ and $X$.

Since there are no mixings, the masses for $\phi_{R}$ and $\phi_{I}$ are simply given by
\begin{equation}
  \label{Eq:AppPhiRMassShiftChaotic}
  \frac{\partial^{2} V}{\partial \phi_{R}^{\, 2}} \simeq \frac{m^{2}}{8 \, 
\sigma_{0}^{\, 3} \, (1 - 2 \, \beta \, \sigma_{0})} - \frac{\lvert W_{\mod}(\sigma_{0}) \rvert^{2}}{4
\sigma_{0}^{\, 3}} \, ,
\end{equation}
and
\begin{equation}
  \label{Eq:AppPhiIMassShiftChaotic}
  \frac{\partial^{2} V}{\partial \phi_{I}^{\, 2}} \simeq \frac{m^{2}}{8 \, 
\sigma_{0}^{3} \, (1 - 2 \, \beta \, \sigma_{0})} \, ,
\end{equation}
respectively. As in the explicit example we discussed in section~\ref{Sec:ExampleVacuumStability}, $\sigma_{0}$ has to satisfy the upper bound
\begin{equation}
  \label{Eq:AppT0ValueBoundShiftChaotic}
  \sigma_{0} < \frac{1}{2 \beta} \, ,
\end{equation}
which is necessary to avoid both a tachyonic mass for $\phi_{I}$ and a wrong sign for the kinetic term of $X$. From the stability condition for $\phi_{R}$, we get an upper bound on the size of $W_{\mod}$ at the minimum in terms of $m$:
\begin{equation}
  \label{Eq:AppWmodBoundVacStabShiftChaotic}
  \lvert W_{\mod}(\sigma_{0}) \rvert^{2} \lesssim \frac{m^{2}}{2 \, (1 - 2 \, \beta \, \sigma_{0})} \approx 2 m^{2} \, ,
\end{equation}
where the last step assumes $\sigma_{0} \approx \sigma_{\inf} \equiv \frac{3}{8 \beta}$. There is again no further constraint from the stability of $x_{R,I}$. Note that this constraint will not be affected by adding the uplifting sector as long as the uplifting term $V_{\up}$, Eq.~\eqref{Eq:AppVuplift}, does not depend on $\Phi$. Furthermore, if any mixings between $T$ and $\Phi$ would be present, the bound is not directly on $W_{\mod}$, but on a particular combination of $W_{\mod}$ and $D_{T} W_{\mod}$ depending on the mixing terms. However, since we consider only setups where $\lvert D_{T} W_{\mod}(\sigma_{0}) \rvert \lesssim \lvert W_{\mod}(\sigma_{0}) \rvert$, the upper bound does not change qualitatively.

\subsection{Corrections to the Inflationary Trajectory}
\label{Sec:AppGeneralWmodCorrectionsInflation}

The presence of the moduli stabilizing sector shifts the inflationary trajectory. Without $V_{\mod}$, it is given by $\sigma = \sigma_{\inf} \equiv \frac{3}{8 \beta}$, $\alpha = 0$,\footnote{See foonote~\ref{Foot:FrozenAxion}.} $X = 0$, $\phi_{R} = 0$ and $\phi_{I} \neq 0$. Upon adding $V_{\mod}$, only $\sigma$ and $x_{I}$ receive shifts since except for the derivative with respect to $\phi_{I}$ the only non-vanishing first derivatives along the would-be inflationary trajectory are
\begin{equation}
  \label{Eq:TShiftInflationTrajectoryShiftChaotic}
   \frac{\partial V}{\partial \sigma} = \left( \frac{D_{T} W_{\mod} \, \overline{W}_{\mod}}{2 \, \sigma^{\, 3}}  +
\frac{D_{T} W_{\mod} \, \overline{W}_{\mod}^{\prime \prime}}{6 \, \sigma} + \text{c.c.} \right) - \frac{2 \lvert D_{T} W_{\mod} \rvert^2}{3 \, \sigma^{\, 2}} -
\frac{3 \lvert W_{\mod}(\sigma_{0}) \rvert^2}{4 \, \sigma^{\, 3} \, \sigma_{0}} \, ,
\end{equation}
and
\begin{equation}
  \frac{\partial V}{\partial x_{I}} = \frac{m \, \phi_{I}}{8 \, \sigma^{\, 3}} \left(
W_{\mod} + \overline{W}_{\mod} \right) \, .
\end{equation}
Note that we have tuned the constant $c_\up$ in $V_\up$ such that it (slighlty over-) compensates the
negative cosmological constant assuming one obtains a supersymmetric AdS vacuum without $V_{\up}$, which yields
\begin{equation}
  \label{Eq:CUpTuningAdSWmod}
  c_{\up} \simeq \frac{3 \lvert W_{\mod}(\sigma_{0}) \rvert^2}{2 \sigma_{0}} \, .
\end{equation}
Thus, the last term in Eq.~\eqref{Eq:TShiftInflationTrajectoryShiftChaotic} is precisely the contribution from $V_{\up}$. If no uplifting is necessary, the corresponding terms in Eq.~\eqref{Eq:TShiftInflationTrajectoryShiftChaotic} and in all of the following equations simply have to be dropped. If the minimum is a non-supersymmetric AdS or Minkowski minimum, $c_{\up}$ is fixed in terms of $W_{\mod}(\sigma_{0})$ and $D_{T} W_{\mod}(\sigma_{0})$ instead of just $W_{\mod}(\sigma_{0})$ and it is straightforward to change to the correct approximate expression for $c_{\up}$ in all the equations.

In the following, we present the results of a perturbative expansion in $W_{\mod}$, $D_{T} W_{\mod}$ and $W_{\mod}^{\prime \prime}$. Note that unless stated otherwise it is implicit that these functions are evaluated at $\sigma = \sigma_{\inf} \equiv \frac{3}{8 \beta}$. To leading order in the expansion, the shifts $\delta x_{I}$ and $\delta \sigma$ are given by
\begin{equation}
  \label{Eq:LeadingOrderDeltaXImInflationShiftChaotic}
  \delta x_{I} \simeq -\frac{\phi_{I} \left( W_{\mod} + \overline{W}_{\mod}
\right)}{m (1 + 32 \gamma \phi_{I}^{2})} \sim \frac{\sigma_{\inf}^{3/2} \, m_{3/2}}{m \phi_{I}} \, ,
\end{equation}
and
\begin{equation} 
\label{Eq:LeadingOrderDeltaTInflationShiftChaotic}
  \begin{split}
  \delta \sigma \simeq \,\, & \frac{\sigma_{\inf}^2 \lvert W_{\mod}(\sigma_{0}) \rvert^2}{4 m^{2}
\sigma_{0} \phi_{I}^{2}} -\frac{\sigma_{\inf} \left(W_{\mod} + \overline{W}_{\mod}
\right)^2}{16 m^{2} \left(1 + 32 \gamma \phi_{I}^2 \right)^2} + \frac{2 \, 
\sigma_{\inf}^3 \, \lvert D_{T} W_{\mod} \rvert^2}{9 m^{2} \phi_{I}^2}
\\ & - \left( \frac{\sigma_{\inf}^{4} \, D_{T}
 W_{\mod} \, \overline{W}_{\mod}^{\prime \prime}}{18 m^{2} \phi_{I}^{2}} +
\text{c.c.} \right) + \left( -\frac{\sigma_{\inf}^{2} \, D_{\bar{T}} \overline{W}_{\mod}}{8 m^{2} (1 +
32 \gamma \phi_{I}^2)} \overline{W}_{\mod}  + \text{c.c.} \right)
\\ & +
  \left( \left(\frac{4}{\phi_{I}^2} -
\frac{3}{1 + 32 \gamma \phi_{I}^2}\right) \frac{\sigma_{\inf}^{2} \, D_{T} W_{\mod}}{24 m^2}
\overline{W}_{\mod} + \text{c.c.} \right)  \, .
  \end{split}
\end{equation}
The first term in $\delta \sigma$ is induced by the uplifting potential $V_{\up}$. Note that
the shift $\delta x_{I}$ is entirely controlled by $W_{\mod}$, i.e. by the gravitino mass during inflation. These
shifts induce some changes to the potential along the $\phi_{I}$-direction, in particular, the
shift of $x_{I}$ away from zero is potentially dangerous. To leading order, the inflaton potential including the shifts is now given by
\begin{equation}
  \label{Eq:GeneralInflationPotentialShiftChaotic}
  V_{\inf} \simeq \frac{m^{2} \phi_{I}^{\, 2}}{4 \, \sigma_{\inf}^{\, 3}} + V_{\mod}(\sigma_{\inf}) - \frac{\phi_{I}^2
\left(W_{\mod} + \overline{W}_{\mod} \right)^2}{16 \, \sigma_{\inf}^{\, 3} \, (1 + 32 \, \gamma \,
\phi_{I}^{\, 2})} \, ,
\end{equation}
where the second term is the pure moduli potential evaluated at $\sigma = \sigma_{\inf}$, i.e. 
\begin{equation}
  \label{Eq:GeneralPureModuliPotential}
  V_{\mod}(\sigma_{\inf}) = \frac{\lvert D_{T} W_{\mod} \rvert^2}{6 \, \sigma_{\inf}} - \frac{3 \lvert W_{\mod} \rvert^2}{8 \, \sigma_{\inf}^{\, 3}} + \frac{3 \lvert W_{\mod}(\sigma_{0}) \rvert^2}{8 \, \sigma_{\inf}^{\, 2} \sigma_{0}} \, ,
\end{equation}
with the last term in Eq.~\eqref{Eq:GeneralPureModuliPotential} coming from the uplifting potential $V_{\up}$, Eq.~\eqref{Eq:AppVuplift}, with $c_{\up}$ tuned as in Eq.~\eqref{Eq:CUpTuningAdSWmod}. The first term in Eq.~\eqref{Eq:GeneralInflationPotentialShiftChaotic} is the standard chaotic inflation potential, which is simply
rescaled by a factor $\sigma_{\inf}^{-3}$ from the $e^{K}$ prefactor in $V_{F}$ and the last term is simply $\sim m^{2} \, \delta x_{I}^{2}$.

Since we must have $\gamma \gtrsim \mathcal{O}(1)$ to ensure that $X$ has a mass
$m_{X} \gtrsim \mathcal{H}$ during inflation, the $\phi_{I}$-dependence of the potential at large
values of $\phi_{I}$ is not significantly affected. Moreover, recall
that from the stability of the vacuum after inflation and with $\sigma_{\inf} \approx
\sigma_{0}$, we must obey the upper bound $\lvert W_{\mod} \rvert \lesssim \sqrt{2} m$, cf. Eq.~\eqref{Eq:AppWmodBoundVacStabShiftChaotic}. Thus, already at this stage, we see that no large corrections should be expected in the regime where our treatment is valid.

In the next step, we have to compute the impact of the extra terms on the inflationary predictions. Since we perform a perturbative expansion, we do not expect to get very large effects, but the bound on $W_{\mod}$ might change. Note that the parameter $m$ has to be redefined from its ``standard'' value, both due to the $\sigma_{\inf}^{\, -3}$ factor in the first term of Eq.~\eqref{Eq:GeneralInflationPotentialShiftChaotic} and due to the extra terms. However, as we will see below, the latter turns out to be irrelevant.



Assuming inflation ends at $\phi_{I} \approx 0$, the number of e-folds $N_{e}$ in terms of the initial value of $\phi_{I}$ at leading order in our perturbative expansion is given by
\begin{equation}
  \label{Eq:NeShiftChaotic}
  \begin{split}
  N_{e} \simeq \,\, & \frac{\phi_{I}^{2}}{4} + \frac{\sigma_{\inf}^{2} \lvert D_{T}
W_{\mod} \rvert^{2}
\ln \phi_{I}}{3 m^{2}} + \frac{3 \, \sigma_{\inf} \lvert W_{\mod}(\sigma_{0}) \rvert^{2} \ln
\phi_{I}}{4 m^{2} \sigma_{0}} \\ & - \left(\frac{1 + (1+ 32 \, \gamma \, \phi_{I}^{2} ) \ln(1 + 32 \, \gamma \, \phi_{I}^{2})}{512 \, \gamma (1 + 32 \, \gamma \, \phi_{I}^{2})} \right) \frac{W_{\mod}^{2} + \overline{W}_{\mod}^{2}}{m^{2}} \\ & + \left( - \frac{1}{\gamma (1 + 32 \, \gamma \, \phi_{I}^{2})} - 192 \ln(\phi_{I}) - \frac{\ln(1 + 32 \, \gamma \, \phi_{I}^{2})}{\gamma} \right) \frac{\lvert W_{\mod} \rvert^{2}}{256 \, m^{2}} \, .
  \end{split}
\end{equation}
There is some additional $\phi_{I}$-dependence, but it is rather weak at
large values of $\phi_{I}$. Thus, we again perform a perturbative analysis to
determine the required initial value $\phi_{I}$ as a function of $N_{e}$, which yields for $N_{e} \gg 1$
\begin{equation}
  \label{Eq:phiNeShiftChaotic}
  \begin{split}
  \phi_{I}(N_{e}) \simeq \,\, & 2 \sqrt{N_{e}} - \frac{3 \sigma_{\inf} \, 
\lvert W_{\mod}(\sigma_{0}) \rvert^{2} \ln(4 N_{e})}{8 m^{2} \sqrt{N_{e}} \sigma_{0}} -
\frac{\sigma_{\inf}^{2} \lvert D_{T} W_{\mod} \rvert^{2} \ln(4 N_{e})}{6 m^{2} \sqrt{N_{e}}}
\\ & + \frac{\left( W_{\mod}^{2} + \overline{W}_{\mod}^{2} \right) \ln(128 \, \gamma \, N_{e})}{512 \, m^{2} \, \sqrt{N_{e}}} + \frac{\left(96 \, \gamma \, \ln(4 N_{e}) + \ln(128 \, \gamma \, N_{e}) \right) \lvert W_{\mod} \rvert^{2}}{256 \, \gamma \, m^{2} \, \sqrt{N_{e}}} \, .
  \end{split}
\end{equation}
Plugging this into the definitions for $\epsilon$ and $\eta$ yields at leading
order
\begin{equation}
  \label{Eq:LeadingOrderEpsilonNeShiftChaotic}
  \begin{split}
  \epsilon \simeq \,\, & \frac{1}{2 N_{e}} + \frac{\sigma_{\inf}^{2} \, \lvert D_{T}
W_{\mod} \rvert^{2} (\ln(4 N_{e})-2)}{12 \, m^{2} \, N_{e}^2} + \frac{3 \, \sigma_{\inf} \lvert W_{\mod}(\sigma_{0}) \rvert^{2}
(\ln(4 N_{e})-2)}{16 \, \sigma_{0} \, m^{2} \, N_{e}^2} 
\\ & - \frac{-2 + 96 \, \gamma (\ln(4 N_{e}) - 2) + \ln(128 \, \gamma \, N_{e}) \lvert W_{\mod} \rvert^{2}}{512 \, \gamma \, m^{2} \, N_{e}^{2}} \\ & - \frac{\left( \ln(128 \, \gamma \, N_{e}) - 2 \right) \left(W_{\mod}^{2} +\overline{W}_{\mod}^{2} \right)}{1024 \, \gamma \, m^{2} \, N_{e}^{2}} \, ,
  \end{split}
\end{equation}
and
\begin{equation}
  \label{Eq:LeadingOrderEtaNeShiftChaotic}
  \begin{split}
  \eta \simeq \,\, & \frac{1}{2 N_{e}} + \frac{\sigma_{\inf}^{2} \, \lvert D_{T}
W_{\mod} \rvert^{2} (\ln(4 N_{e}) - 1)}{12 \, m^{2} \, N_{e}^2} + \frac{\sigma_{\inf} \lvert W_{\mod}(\sigma_{0}) \rvert^{2}
(\ln(4 N_{e}) - 1)}{16 \, \sigma_{0} \, m^{2} \, N_{e}^2} 
\\ & - \frac{-1 + 96 \, \gamma (\ln(4 N_{e}) - 2) + \ln(128 \, \gamma \, N_{e}) \lvert W_{\mod} \rvert^{2}}{512 \, \gamma \, m^{2} \, N_{e}^{2}} \\ & - \frac{\left( \ln(128 \, \gamma \, N_{e}) - 1 \right) \left(W_{\mod}^{2} + \overline{W}_{\mod}^{2} \right)}{1024 \, \gamma \, m^{2} \, N_{e}^{2}} \, ,
  \end{split}
\end{equation}
respectively. Note that all the corrections start at order $N_{e}^{-2}$ (up to some logarithms):
They are suppressed with respect to the leading contribution by $W_{\mod}^{2} / W_{X}^{2}$ etc. since $m^{2} N_{e} \sim m^{2} \phi_{I}^{2} = W_{X}^{2}$.

Before we continue, we have to fix the
parameter $m$ by matching the prediction to the observed amplitude of the scalar power spectrum $\mathcal{P}_{\mathcal{R}}^{1/2}$. We find
\begin{equation}
  \label{Eq:LeadingOrderScalarAmplitudeShiftChaotic}
  \begin{split}
  \mathcal{P}_{\mathcal{R}}^{1/2} \simeq \,\, & \frac{m N_{e}}{2 \sqrt{3} \pi
\sigma_{\inf}^{3/2}} \Bigg(1 - \frac{\sigma_{\inf}^2 \lvert D_{T} W_{\mod} \rvert^{2} (2 \ln(4 N_{e}) - 3)}{12 \, m^{2} \, N_{e}}  - \frac{3 \, \sigma_{\inf} \lvert W_{\mod}(\sigma_{0}) \rvert^{2} (2 \ln(4 N_{e}) - 3)}{16 m^{2} N_{e}
\sigma_{0}} \\ & + \frac{\left(2 \ln(128 \, \gamma \, N_{e}) - 3 \right) \left(W_{\mod}^{2} +\overline{W}_{\mod}^{2} \right)}{1024 \, \gamma \, m^{2} \, N_{e}} \\ & + \frac{\left( -3 + 96 \, \gamma \, (2 \ln(4 N_{e}) - 3 )  + 2 \ln(128 \, \gamma \, N_{e}) \right) \lvert W_{\mod} \rvert^{2}}{512 \, \gamma \, m^{2} \, N_{e}} \Bigg) \, .
  \end{split}
\end{equation}
Thus, except for the factor $\sigma_{\inf}^{3/2}$, the mass parameter $m$ needs to be redefined only at second order in
$W_{\mod}$ and $D_{T} W_{\mod}$. Consequently, this
affects the above expressions for $\epsilon$ and $\eta$ only at higher orders
and we drop these corrections in the following. What matters, however, is the
rescaling of the mass parameter $m$ by $\sigma_{\inf}^{3/2}$ compared to the standard chaotic inflation scenario.

Now we can calculate the scalar spectral index $n_{s} = 1 - 6 \, \epsilon + 2 \, \eta$ and the tensor-to-scalar ratio $r = 16 \, \epsilon$, for which we find
\begin{equation}
  \label{Eq:LeadingOrderScalarIndexShiftChaotic}
  \begin{split} 
  n_s - 1 \simeq \, & -\frac{2}{N_{e}} - \frac{\sigma_{\inf}^{2} \lvert D_{T} W_{\mod} \rvert^{2}
(2 \ln(4 N_{e}) - 5)}{6 m^{2} N_{e}^2}  - \frac{3 \, \sigma_{\inf} \lvert W_{\mod}(\sigma_{0}) \rvert^{2}
(2 \ln(4 N_{e}) - 5)}{8 m^{2} N_{e}^{2} \sigma_{0}}
\\ & + \frac{ \left( - 5 + 96 \, \gamma (2 \ln(4 N_{e}) - 5) + 2 \ln(128 \, \gamma \, N_{e})  \right) \lvert W_{\mod} \rvert^{2}}{256 \, \gamma \, m^{2} \, N_{e}^{2}}
\\ & + \frac{ \left(2 \ln(128 \, \gamma \, N_{e}) - 5  \right)  \left(W_{\mod}^{2} + \overline{W}_{\mod}^{2} \right)}{512 \, \gamma \, m^{2} \, N_{e}^{2}}  \, ,
  \end{split}
\end{equation}and
\begin{equation}
  \label{Eq:LeadingOrderTensorToScalarRatioShiftChaotic}
  \begin{split}
  r \simeq \, & \frac{8}{N_{e}} + \frac{4 \, \sigma_{\inf}^{2} \, \lvert D_{T}
W_{\mod} \rvert^{2} (\ln(4 N_{e})-2)}{3 \, m^{2} \, N_{e}^2} + \frac{3 \, \sigma_{\inf} \lvert W_{\mod}(\sigma_{0}) \rvert^{2}
(\ln(4 N_{e})-2)}{\sigma_{0} \, m^{2} \, N_{e}^2} 
\\ & - \frac{-2 + 96 \, \gamma (\ln(4 N_{e}) - 2) + \ln(128 \, \gamma \, N_{e}) \lvert W_{\mod} \rvert^{2}}{32 \, \gamma \, m^{2} \, N_{e}^{2}} \\ & - \frac{\left( \ln(128 \, \gamma \, N_{e}) - 2 \right) \left(W_{\mod}^{2} + \overline{W}_{\mod}^{2} \right)}{64 \, \gamma \, m^{2} \, N_{e}^{2}} \, ,  \end{split}
\end{equation}
respectively. Obviously, the corrections with respect to the leading term are small. They also appear at a higher order in the large $N_{e}$ expansion since the corrections are suppressed by $W_{X}^{2} = m^{2} \Phi^{2} \sim m^{2} N_{e}$ with respect to the leading contribution. Low-energy supersymmetry and high-scale inflation correspond to $D_{T} W_{\mod}$ and $W_{\mod}$ being parametrically small compared to the F-term $D_{X} W$. Consequently, all the corrections are parametrically small as well since we have to require $\sigma_{0} \approx \sigma_{\inf}$ to avoid the cosmological moduli problem and we assume that $D_{T} W_{\mod}$ and $W_{\mod}$ do not vary too strongly between $\sigma_{0}$ and $\sigma_{\inf}$.

\end{document}